\documentclass[preprint,12pt]{aastex}
\usepackage{epsfig, amsmath, amsfonts, graphicx, amssymb}

\def\gtsima{$\; \buildrel > \over \sim \;$}
\def\gtsim{\lower.5ex\hbox{\gtsima}}

\def\beq{ \begin{equation} }
\def\eeq{ \end{equation} }

\begin{document}


\title{\bf \Large On the high frequency spectrum of a classical accretion disc}

\author{ Steven A. Balbus\altaffilmark{1}}

\altaffiltext{1}{Dept. Physics, University of Oxford, Keble Rd.,
Oxford, UK OX1 3RH
  \texttt{steven.balbus@lra.ens.fr}}

\begin{abstract}
We derive simple and explicit expressions for the high frequency
spectrum of a classical accretion disc.   Both stress-free and
finite stress inner boundaries are considered.  A classical accretion
disc spectrum with a stress-free inner boundary departs from a Wien
spectrum at large $\nu$, scaling as $\nu^{2.5}$ (as opposed to
$\nu^3$) times the usual exponential cut-off. If there is finite
stress at the inner disc boundary, the maximum disc temperature
generally occurs at this edge, even at relatively modest values of
the stress.  In this case, the high frequency spectrum is proportional
to $\nu^2$ times the exponential cut-off.  If the temperature maximum
is a local hot spot, instead of an axisymmetric ring, then an
interior maximum produces a $\nu^2$ prefactor while an edge maximum
yields $\nu^{1.5}$.  Because of beaming effects, these latter
findings should pertain to a classical relativistic disc.   The
asymptotics are in general robust and independent of the detailed
temperature profile, provided only that the liberated free energy
of differential rotation is dissipated locally, and may prove useful
beyond the strict domain of classical disc theory.  As observations
continue to improve with time, our findings suggest the possibility
of using the high energy spectral component of black hole candidates
as a signature prediction of classical theory, as well as an
diagnostic of the stress at the inner regions of an accretion disc.

\end{abstract}

\maketitle

\section {Introduction}

Well into its fifth decade (e.g. Lynden-Bell 1969, Shakura \& Sunyaev 1973),
classical thin disc accretion theory has withstood the test of time reasonably
intact.  While it is clear that accretion is considerably more complex
than allowed for in classical disc theory (hereafter CDT), there are many examples in
which its computed disc spectrum, a superposition of blackbody rings
over the surface of the disc, is realised with some fidelity in
observations.  Even when there are temporal changes in the spectra,
there is generally a rather well-defined thin disc state to be 
found (e.g. Davis et al. 2005).  A detailed theoretical
examination of the disc spectrum is both useful and revealing.

In its classical and simplest formulation (Lynden-Bell 1969, Pringle
1981, Frank, King, \& Raine 2002), the computation of the disc
spectrum involves an integral that must in general be evaluated
numerically.   In this {\em Letter}, we show that the integral in
question may, however, be performed analytically in the astrophysically
relevant limit $E_\gamma\gg kT_{max}$, where $E_\gamma$ is the
photon energy, $k$ the Boltzmann constant, and $T_{max}$ the maximum
disc temperature.  The resulting spectra are similar to, but depart
from, a Wien spectrum.  The  computed frequency dependence is both
robust and insensitive to the precise temperature profile.  The
form of the departure from a Wien spectrum changes (nearly
discontinuously) beyond a moderate threshold value of the stress
at the inner disc edge.  It is this result that may prove useful
to the analysis of the dynamical conditions at the inner disc edge.
The calculational technique, however, should find application under
any conditions in which the disc has a well-defined temperature
maximum.   In principle, the effect of several isolated maxima may
be superposed, but in general the spectrum will be dominated at
high $\nu$ by the highest temperature peak.

The calculations are presented in the next section, which is followed by
a brief discussion of our results.  Both relativistic and nonrelativistic 
discs are considered.

\section{Disc spectrum}

\subsection {Preliminaries}

In nonrelativistic CDT, the emitted flux per unit frequency $\nu$ from a thin disc
is given by the expression (e.g. Frank, King, \& Raine 2002):
\beq\label{fnu}
F_\nu = {4\pi h\nu^3\cos i\over c^2 r^2} \int^\infty_{R_*} {R\, dR\over
\exp[h\nu/kT(R)] - 1}
\eeq
Where $h$ is Planck's constant, $c$ the speed of light, $r$ the distance to the source,
$i$ the inclination angle (from a face-on orientation), $R_*$ the inner disc radius,
and $T(R)$ the temperature profile.  The integral is over all radii $R$ and formally extends
to infinity.  

The time-steady surface 
temperature profile of a thin Keplerian disc is given by (Pringle 1981, Balbus
\& Hawley 1998):
\beq\label{t1}
T^4(R) = {3GM{\dot M}\over 8\pi R^3 \sigma} \left[ 1 - \left(R_*\over R\right)^{1/2}\right]
\eeq
where $G$ is gravitational constant, $\sigma$ the Stefan-Boltzmann constant,
$M$ the central mass, $\dot M$ the steady mass
accretion rate.  This result is calculated by assuming that the stress vanishes at
the inner edge $R=R_*$.  But $R_*$ might equally well be regarded as an integration constant 
(arising from angular momentum flux conservation) whose value 
is determined by specifying the stress at some radius.  To avoid confusion, we will
write 
\beq\label{t2}
T^4(R) = {3GM{\dot M}\over 8\pi R^3 \sigma} \left[ 1 - \left(R_0\over R\right)^{1/2}\right]
\eeq
where $R_0$ is an arbitrary constant with dimensions of a length
whose value we leave unspecified.  The choice $R_0=R_*$
corresponds to what we shall view as the special case of the stress vanishing at the inner
edge of the disc.  Other notational conventions we will use are
\beq
\Omega^2 = {GM\over R^3}, 
\eeq
for the Keplerian angular velocity (s$^{-1}$), 
$\Sigma$ for the height-integrated
disc surface density, and the characteristic temperatures
\beq\label{t0}
T_0^4 = {3GM{\dot M}\over 8\pi R_0^3 \sigma},\quad
T_*^4 = {3GM{\dot M}\over 8\pi R_*^3 \sigma}.
\eeq
Care should be taken to distinguish $T_*$ and $T(R_*)$, which are
quite distinct.  The dominant $R\phi$ component
of the density-weigted velocity stress tensor (dimensions of velocity$^2$) is denoted
$W_{R\phi}$.

The temperature reaches a maximum, $T_{max}$, when $R_{max}=(49/36)R_0$
(Kubota et al. 1998).  This is near the inner
edge if $R_0$ also is, but there may be no temperature maximum if $R_0\ll R_*$.  This
corresponds to the case of 
a significant stress at the inner edge.  ``Significant'' means a value
close to the local ${\dot M}\Omega/(2\pi\Sigma)$ (Balbus \& Hawley 1998).  
We will show that the two cases of an interior temperature
maximum and a boundary maxium lead to distinct observational signatures.

\subsection {Finite stress modification}

The condition of angular momentum conservation may
quite generally be written (Balbus \& Hawley 1998)
\beq\label{WRP}
-{{\dot M}R^2\Omega\over 2\pi} +\Sigma R^2 W_{R\phi} = {\rm constant} =
-{{\dot M}R_*^2\Omega_*\over 2\pi} +\Sigma_* R_*^2 W  
\eeq
where $W$ is the selected value of $W_{R\phi}$ at $R=R_*$.
(Both $\Sigma_*$ and $\Omega_*$ are evaluated at $R=R_*$.)   This
value of $W_{R\phi}$ 
is to be distinguished from a fiducial characteristic value
\beq
W_{char}  = {{\dot M} \Omega_*\over 2\pi \Sigma_*},
\eeq
which will appear (as a normalization for $W$) in the equations.

If we now solve equation (\ref{WRP}) for $W_{R\phi}$, we obtain
\beq
W_{R\phi} = {{\dot M}\Omega\over 2\pi \Sigma} \left[ 1 - \left(R_0\over
R\right)^{1/2} \right]
\eeq
where 
\beq\label{R0}
R_0 = R_* \left( 1- {W\over W_{char}}\right)^2
\eeq
Equating the energy radiated by (each side of) the disc to the energy extracted from
the differential rotation, we have (Balbus \& Hawley 1998):
\beq\label{ss}
 2 \sigma T^4 = - \Sigma W_{R\phi} {d\Omega\over d\ln R} = 
- {{\dot M}\over 4\pi } {d\Omega^2\over d\ln R}\left[ 1 - \left(R_0\over
R\right)^{1/2} \right],
\eeq
leading immediately to (\ref{t2}). 
Equation (\ref{R0}) tells us precisely how the $R_0$ constant is related
to the imposed stress $W$ and inner boundary $R_*$, and is for that reason very useful.  
As noted, the formal location of the temperature maximum from (\ref{t2}) is
$R_{max}=(49/36)R_0$.  The question is, at what value of $W$ does this radius
move from within the disc proper to the inner edge?
Setting $R_{max}=R_*$ and using (\ref{R0}) leads to 
\beq\label{wsmall}
W={W_{char}\over 7}
\eeq
Thus, when
$W$ exceeds $0.1429W_{char}$, the temperature maximum lies on the inner disc
boundary, and when the stress drops below this, the temperature maximum moves off the
boundary to within the disc inerior.
As we shall now see, the location of the temperature maximum makes a signficant
difference to the emitted spectrum.   

\subsection {Large $\nu$ limit of $F_\nu$.}

\subsubsection{Small stress: interior $T_{max}$}

Consider the integral
\beq\label{II}
I = \int^\infty_{R_*} {R\, dR\over \exp[h\nu/kT(R)] - 1}
\eeq
with $T(R)$ given by (\ref{t2}).
When $h\nu\gg kT_{max}$ we may safely ignore the $-1$ in the denominator
across the entire domain of integration,
since it leads only to exponentially small corrections.  Thus,
\beq\label{I}
I = \int^\infty_{R_*} {R \exp[-h\nu/kT(R)]}\, dR
\eeq
The function $\beta \equiv 1/kT(R)$ has a sharp {\em minimum} at
$R=R_{max}$, which renders the integral an ideal candidate for
an asymptotic expansion based on Laplace's method (Bender \& Orszag 1978).
Under these conditions, the entire contribution to the integral comes
from a small region near $R=R_{max}$.  Expanding $\beta$ and remembering
the first derivative vanishes at $R=R_{max}$:
\beq\label{beta1}
\beta = {1\over kT_{max}} + \beta''_{min}{(R-R_{max})^2\over 2} + ...
\eeq
where 
\beq
\beta''_{min} = {d^2\ \over dR^2} \left[1\over kT(R)\right]_{R=R_{max}}
\eeq
The integral (\ref {I}) transforms to 
\beq
I \simeq  R_{max}\exp(-h\nu/kT_{max})\,  \int^\infty_{-\infty} \exp (-h\nu \beta''_{min}x^2/2)\, dx
\eeq
where $x=R-R_{max}$, and we set $R=R_{max}$ since only this neighborhood contributes.
Extending the limits of integration introduces only exponentially small
corrections.   Hence,
\beq\label{ias}
I \simeq  R_{max} \left(2\pi \over \beta''_{min}h\nu\right)^{1/2}\> \exp(-h\nu/kT_{max})\, 
\eeq

For the distribution (\ref{t2}), $\beta''_{min}$ works out to 
\beq
R_0^2 \beta''_{min} = 3^{7/2}/(7^{5/4}\times 2^{1/2}\times kT_0) = 2.90426/(kT_0)
\eeq
This leads to an emission spectrum
\beq\label{f1}
F_\nu = 2.002 {4\pi\cos i\over c^2}\,   (hk T_0)^{1/2} \,
{R_0^2\over r^2}\>  \nu^{5/2} \exp(-h\nu/kT_{max})
\eeq
where $T_{max} = 0.487871 T_0$.  The solution for a classical zero-stress inner boundary
is obtained by setting by
using the inner edge of the disc $R_*$ for $R_0$ and $T_*$ for $T_0$.   
The key point is that the frequency dependence differs from a Wien spectrum
by a factor of $\nu^{-1/2}$.

\begin{figure} [h]
\includegraphics[width=12cm] {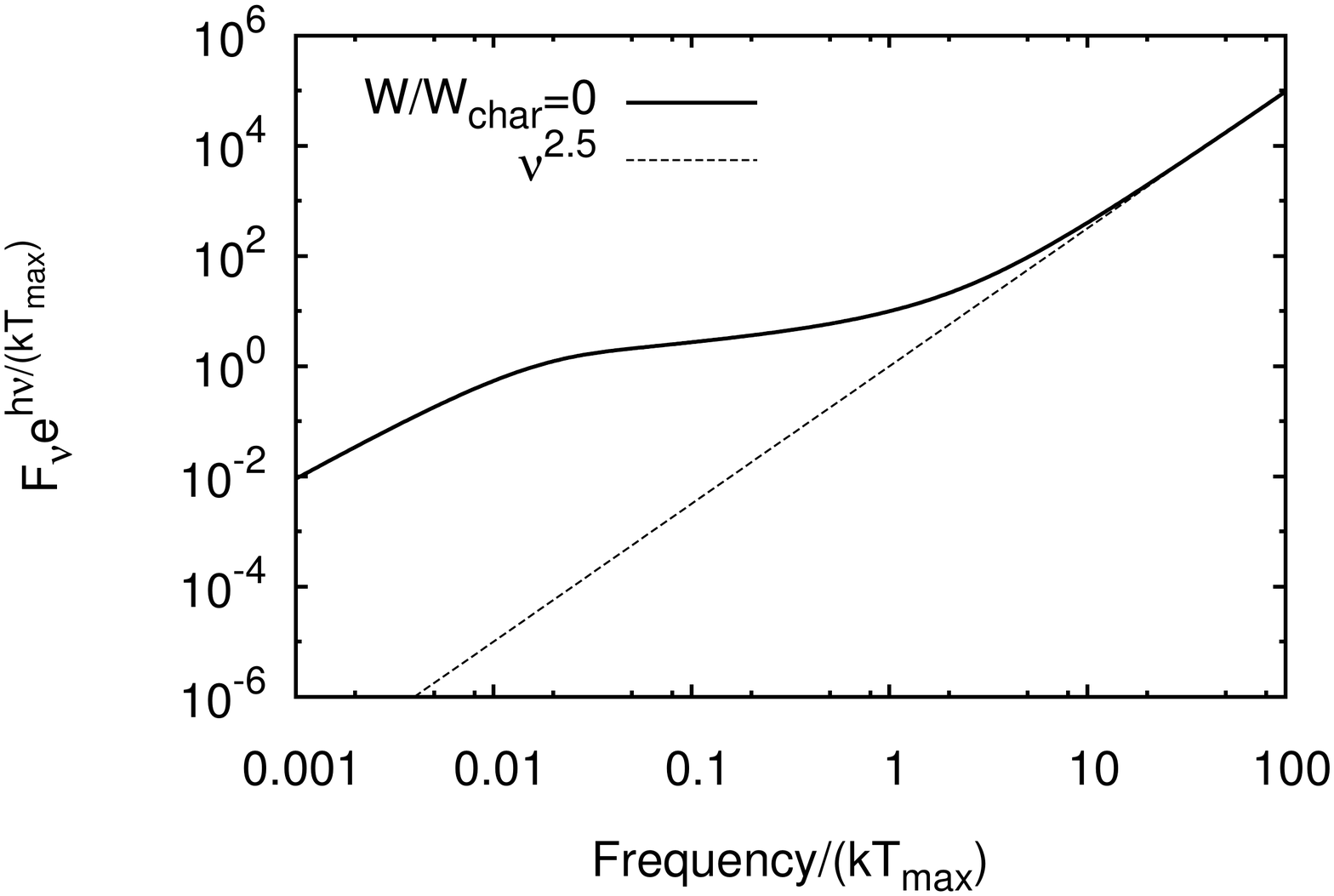}
\caption {{\small Plot comparing a $ F_\nu e^{h\nu/kT_{max}}$, renormalised for display, with the
large $\nu$  asymptotic result
for the case of vanishing stress. Solid line is from numerical evaluation;
dotted line is large $\nu$ asymptotic form (eq. [\ref{ias}]).}}
\end{figure}

\subsubsection{Moderate-to-large stress: boundary $T_{max}$}

If the stress $W$ exceeds $W_{char}/7$, the temperature maximum moves to the boundary.  In
that case, equations (\ref{t2}) and (\ref{R0}) may be combined to yield the
temperature at the inner disc edge, $T(R_*)$:
\beq
T(R_*) = T_* w^{1/4}, \quad {\rm with\ } w\equiv W/W_{char}.
\eeq
Another quantity of interest we shall require is the temperature gradient at $R=R_*$.
This is most conveniently expressed in the form
\beq
\left(d\ln R\over d\ln\beta\right)_{R=R_*} = 
 - \left(d\ln R\over d\ln T\right)_{R=R_*}  
={w\over 2(7w-1)}.
\eeq

The high frequency behaviour of (\ref{I}) is obtained by a simple integration by
parts.   Now the first derivative $\beta'$ term dominates and one finds
\beq\label{ias2}
I \simeq \left(R_*\over h\nu\right) \left(\exp[- h\nu/kT(R_*)]\over \beta'(R_*)\right)
=  R_*^2 \left(kT_*\over h\nu\right) {w^{5/4}\over 2(7w-1)}
\exp(-h\nu/[kT_* w^{1/4}]),
\eeq
where 
\beq
\beta'_* = - {1\over kT_*^2}\left(dT\over dR\right)_{R=R_*}.
\eeq
This gives a spectral flux of 
\beq\label{f2}
F_\nu  = \left(R_*\over r\right)^2
 {2\pi\cos i\over c^2} (kT_*) {w^{5/4}\over 7w-1}
\nu^2\, \exp(-h\nu/[kT_*w^{1/4}])
\eeq
This is a less steep frequency dependence than is present in (\ref{f1}).   The
case of small stress and an interior maximum corresponds to a flatter region
of high temperature, and a correspondingly larger disc area is able to contribute.
A boundary maximum is more steeply cut-off as one moves outward, and less of the disc
is able to contribute, reducing the high $\nu$ emission relative to the interior
maximum case.

Figures (1) and (2) show two representative examples illustrating the region
of validity of our approximation.   Deviations from the leading asymptotic behavior
are expected to be $O(kT_{max}/h\nu)$ in both cases (Bender \& Orszag 1978).  
Quantitatively, the agreement between asymptotic and exact integrals is 
excellent when $h\nu/kT_{max} \ga 5kT_{max}$. 
When the ratio is $10$, the results are indistinguishable, at which point the
flux is about $5\times 10^{-3}$ of its peak value.  

\begin{figure}[h]
\includegraphics[width=12cm] {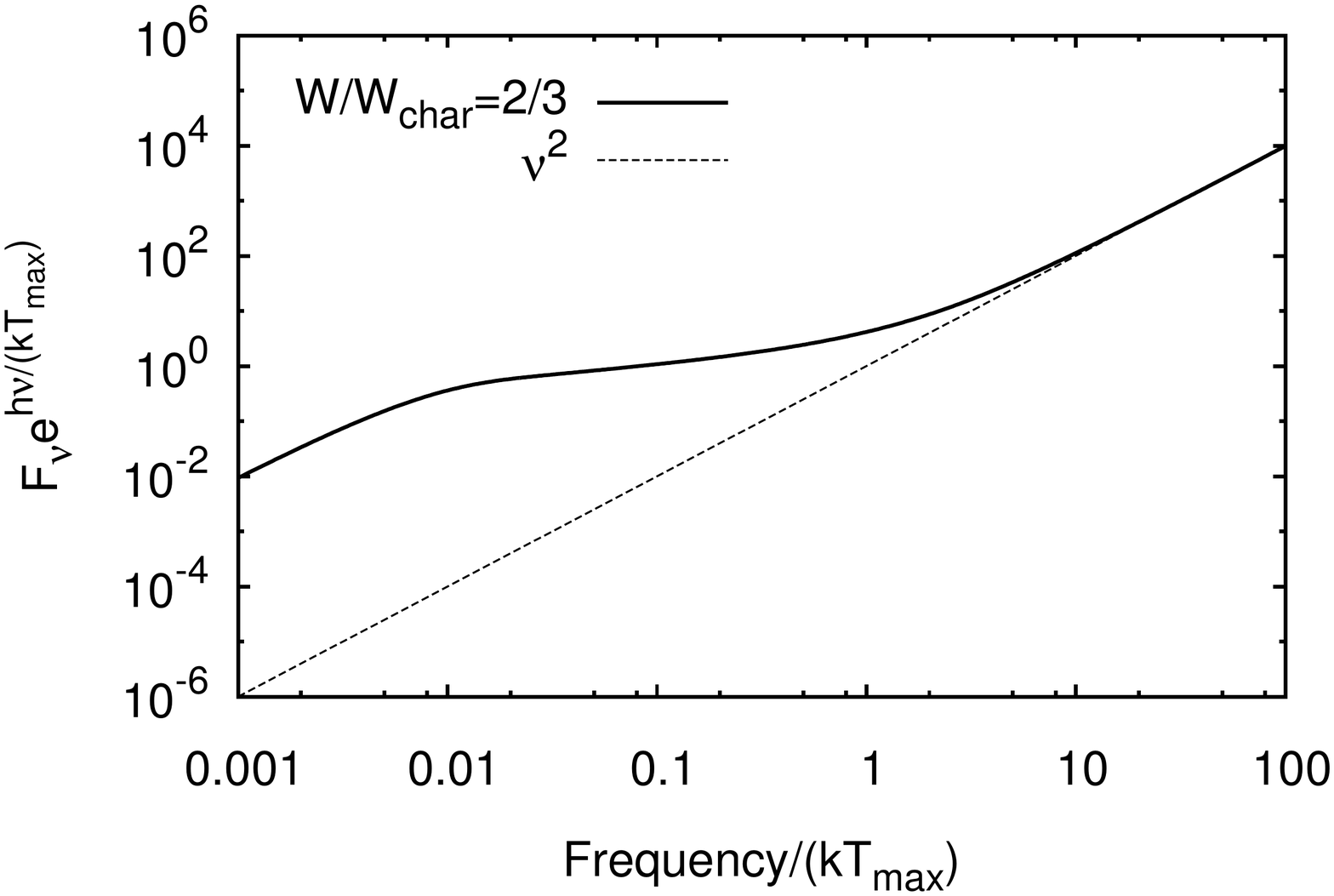}
\caption {{\small As in figure (1) for the finite stress case $W=(2/3)W_{char}$.}}
\end{figure}

\subsection {A localised hot spot}

The axisymmetric form of the emission integral (\ref{II}) of classical disc 
theory is not preserved when relativistic physics is included.  The most
important deviation is due to beaming from the portion of the
disk approaching the observer.  To extract the asymptotic behaviour of the
spectrum, however, we need not explicitly invoke the full machinery of general relativity.
It will suffice to note that the emission integral will be of the form
\beq
I = \nu^ 3 \int_{\cal S}\  {F(s_1, s_2)\, dS\over \exp[h\nu /kT(s_1, s_2)] -1}
\eeq
where ${\cal S}$ is the effective ``working surface,'' $s_1$ and $s_2$ surface
coordinates, $F$ an unspecified function of coordinates (but, importantly, {\it not} $\nu$)
and $dS$ an area element.   If $T$ now has a maximum within the disc interior
localised at some particular location $(s_1, s_2)$, the high $\nu$ contribution
is dominated by a small two-dimensional neighbourhood of this point.  Even
if the disc shape is globally complicated, locally it can always be represented
as a flat plane with Cartesian $(x,y)$ coordinates.  

Near the temperature maximum $T_{max}$ we proceed as in subsection 2.3.1,
expanding the $\beta$ function around the coordinates $x_{max}, y_{max}$:
\beq
\beta = {1\over kT_{max}} +\beta_{xx}{(x-x_{max})^2\over 2} +\beta_{yy}
{(y-y_{max})^2\over 2} + ...
\eeq
where the subscript $x$ or $y$ denotes partial differentiation with the 
other variable held fixed.  The first order partial derivatives vanish at 
the maximum, as does the mixed derivative $\beta_{xy}$ for the proper choice
of coordinates. 
The second order (nonmixed) derivatives are understood 
to be evaluated at $x_{max}, y_{max}$.  Exactly the same reasoning as before
leads to the emission (double) integral
\beq
I = \nu^3 F_{max} \exp[-h\nu/kT_{max}]\, \int_{-\infty}^{\infty} 
\int_{-\infty}^{\infty} 
\exp(-h\nu\beta_{xx}x^2/2)
\exp(-h\nu\beta_{yy}y^2/2) dx\, dy
\eeq
where $F_{max}$ is $F$ evaluated at $x_{max}, y_{max}$.   This yields
\beq
I =  {2\pi F_{max} \over  h (\beta_{xx}\beta_{yy})^{1/2}} \ \nu^2 \exp[-h\nu/kT_{max}]
\eeq
The frequency dependence is exactly that of a classical axisymmetric disc with a
temperature maximum at the inner boundary.   On the other hand, for a disc
with a localised hot spot at the inner boundary, exactly the same techniques
we have been using show that the large $\nu$ asymptotic form is 
\beq
I \sim \nu^{3/2} \exp(-h\nu/kT_{max}) \quad{\rm (Hot\ spot\ on\ inner\ boundary.)}
\eeq


\section {Discussion}

We have shown that at large photon energies, CDT yields
a mathematically simple---and possibly
observationally interesting---difference in the frequency
dependence evinced by a thin disc and a true Wien spectrum. 
For a Novikov-Thorse relativistic disk, the difference
is yet more pronounced.
Of greater astrophysical significance, perhaps, is our finding
that a spectral changes 
of comparable magnitude and simplicity occur in going from
zero to moderate stress at the disc's inner boundary.
This arises because, unless the stress is very small 
(cf eq.[\ref{wsmall}]), the maximum
disc temperature is reached on this inner boundary\footnote{This is
true provided the stress follows the precepts of CDT and
causes local dissipational heating
as described by equation (\ref{ss}). Whether 
magnetic stresses behave in this manner is currently being investigated.  (I thank J. Krolik
for drawing my attention to this important point.)}.  By constrast, a
zero stress constraint always results in a temperature peak within the
disc interior.  These different locations of the maxima cause measurably
different high energy spectra: a $\nu^{5/2}$ ($\nu^2$)
power law multiplying an exponential for the case of an interior (edge) maximum. 
(Recall that a Wien spectrum has a $\nu^3$ power law prefactor.)  
For a relativistic disc, or any other disc in which the maximum is a localised hot spot,
the scalings are $\nu^2$ ($\nu^{3/2}$).
These findings stand on their own, but since
the question of the presence
or absence of an inner stress, which is likely to be
magnetic in origin (e.g. Agol \& Krolik 2000),
is a lively and contested issue
(Beckwith, Hawley, \& Krolik 2008), the current results may not be devoid of 
practical significance.

At the very least, our findings are useful benchmarks for numerical calculations of
disc spectra.  Real discs, onthe other hand,
live in messy accretion environments, often with many 
spectral components.  Principal sources of confusion at high
frequencies include emission from a hot corona and Comptonisation of soft photons.
Even here, there is some utility in knowing the precise frequency dependence
of a thermal disc, if only as a baseline from which to mark differences. 
Moreover, there are discs with minimal coronal components, 
and the models studied by Shimura \& Takahara (1995; see also 
the discussion of Davis et al. 2005) indicate that the effects of Comptonisation can
be well-modelled by replacing the Planck function
in equation (\ref{fnu}) by a ``dilute blackbody'' form. This modification would not
change the frequency dependence of our formulae.   Asymptotic expansions
in the high frequency limit of spectral integrals,
together with data that promises to be ever more accurate, will 
both strengthen and deepen our understanding of compact X-ray sources.
Development and observational applications of these findings,
as well as detailed comparisons with relativistic disc 
models (Novikov \& Thorne 1973), are currently being pursued.

\section*{Acknowledgements}
It is a pleasure to thank Shane Davis, Julian Krolik, and Chris Done for 
extended correspondence and
important advice.   I would also like to acknowledge detailed conversations with 
Omer Blaes, Mari Kolehmainen, 
Will Potter (who also kindly prepared figures 1 and 2), and helpful comments
from an anonymous referee.  
Support from the Royal Society in the form of a Wolfson Research Merit Award 
is gratefully acknowledged.



\begin{thebibliography}{99}

\bibitem[\protect\citeauthoryear{Agol \& Krolik}{2000}]{ak00} Agol, E.,
\& Krolik, J. H. 2000, ApJ, 528, 161

\bibitem[\protect\citeauthoryear{Balbus \& Hawley}{1998}]{bh98} Balbus, S. A., \& Hawley, J. F. 
1998, Rev. Mod. Phys., 70, 1

\bibitem[\protect\citeauthoryear{Beckwith et al}{2008}]{bhk08} Beckwith, K, Hawley, J. F., \&
Krolik, J. H. 2008, MNRAS, 390, 21

\bibitem[\protect\citeauthoryear{Bender \& Orszag}{1978}]{bo78}  Bender, C. M., \& Orszag, S. A.,
1978, Advanced Mathematical Methods for Scientists and Engineers (McGraw-Hill: New York)

\bibitem[\protect\citeauthoryear{Davis et al.}{2005}]{de05} Davis, S. W., Blaes, O. M., Hubeny, I.,
\& Turner, N. J. 2005, ApJ, 621, 372

\bibitem[\protect\citeauthoryear{Frank et al}{2002}]{fkr2} Frank, J., King, A., \& Raine, D. 2002,
Accretion Power in Astrophysics (Cambridge University Press: Cambridge)

\bibitem[\protect\citeauthoryear{Kubota et al}{1998}]{ke98}  
Kubota, Aya, Tanaka, Y., Makishima, K., Ueda, Y., Dotani, T., Inoue, H., \& Yamaoka, K. 1998,
PASJ, 50, 667

\bibitem[\protect\citeauthoryear{Lynden-Bell}{1969}]{l69}  Lynden-Bell, D. 1969, Nature, 223, 690

\bibitem[\protect\citeauthoryear{Novikov \& Thorne}{1973}]{nt73} Novikov, I. D., \& Thorne, K.
1973, in Black Holes---Les Astres Occlus, ed. C. De Witt (Gordon \& Breach: New York), p. 346.

\bibitem[\protect\citeauthoryear{Pringle}{1981}]{p81}  Pringle, J. E. 1981, ARAA, 19, 137      

\bibitem[\protect\citeauthoryear{Shakura \& Sunyaev}{1973}]{ss73} Shakura, N. I., \&
Sunyaev, R. A. 1973, A\&A,24, 337       

\bibitem[\protect\citeauthoryear{Shimura \& Takahara}{1995}]{sh95} Shimura, T., \& Takahara, F. 1995,
ApJ, 445, 780
\end{thebibliography}
\end{document}